\documentclass[11pt,a4paper]{article}
\usepackage[utf8]{inputenc}
\usepackage[T1]{fontenc}
\usepackage{lmodern}
\usepackage[margin=2.5cm]{geometry}
\usepackage{amsmath,amssymb}
\usepackage{graphicx}
\usepackage{rotating}
\usepackage{setspace}
\usepackage[font=small,labelfont=bf]{caption}
\usepackage[hidelinks]{hyperref}
\usepackage{xurl}
\usepackage{xcolor}

\graphicspath{{figs/}}
\hypersetup{
  pdftitle={Original Insights Into Rock Slope Damage Processes Until Collapse From Passive Seismic Monitoring},
  pdfauthor={P. Bottelin, L. Baillet},
  pdfsubject={Preprint},
  pdfkeywords={rock slope, passive seismic monitoring, resonance frequency, damage, rockfall, early warning}
}

\newcommand{\fz}{f_0}
\newcommand{\dfz}{\Delta f_0}

\begin{document}

\begin{center}
\fbox{\parbox{0.92\textwidth}{\small\setstretch{1.1}
\textbf{Preprint.} This is the authors' submitted version of the article published in
\emph{Geophysical Research Letters}, 51 (2024), e2024GL109139.
The final version is available at \url{https://doi.org/10.1029/2024GL109139}.
Supporting Information is available with the published version.
This version may differ from the published version.}}
\end{center}

\vspace{1em}
\begin{center}
{\Large\bfseries Original Insights Into Rock Slope Damage Processes Until Collapse From Passive Seismic Monitoring\par}
\vspace{1.2em}
{\large P.~Bottelin$^{1,*}$ and L.~Baillet$^{2}$\par}
\vspace{0.8em}
{\small\setstretch{1.1}
$^{1}$Société Alpine de Géotechnique (SAGE Ingénierie), Gières, France\\
$^{2}$ISTerre, Université Grenoble Alpes, Université Savoie Mont Blanc, CNRS, IRD, Université Gustave Eiffel, Grenoble, France\\
$^{*}$Corresponding author: p.bottelin@sage-ingenierie.com\par}
\end{center}

\section*{Key Points}
\begin{itemize}
\item We observed a 24\% decrease in rock slope fundamental frequency during the 6 months preceding a $\sim$14,000~m$^3$ collapse
\item New insights into slope degradation processes revealed the alternate control of two damage regimes with contrasting rate
\item Results revealed slope damage phases that would not have been suspected on the basis of slope deformation monitoring alone
\end{itemize}

\section*{Abstract}
We performed a passive seismic monitoring of the La Praz $\sim$14,000~m$^3$ unstable slope (French Alps) spanning over 10 years. During the last 6 months prior to collapse, we detected a clear 24\% decrease in the slope's fundamental resonance frequency, $\fz$, caused by a reduction in overall rock mass stiffness. The combined study of $\fz$ and slope deformation suggested the alternating importance of sudden brittle failure processes versus more ductile phases with possible sliding. Seismic monitoring revealed slope damage that remained ambiguous or undetected with ground surface deformation monitoring, and highlighted critical periods with intense damage. Only some of these critical damage periods could be related to clear external forcing factors such as intense rainfall episodes. These new insights into rock slope's structural condition at depth represent an asset for future monitoring systems. Surface deformation and passive seismic stiffness tracking combined could reveal active slopes with ongoing damage processes.

\section*{Plain Language Summary}
Forecasting the time of rockfalls is of critical importance for risk mitigation operators in order to preserve the safety of persons and the integrity of infrastructure. Most monitoring systems are based on ground deformation measurements, which may fail when the surface motion does not accurately reflect changes in slope stiffness with time. In this work, we used a seismic sensor that passively recorded ground vibrations on top of a $\sim$14,000~m$^3$ unstable slope. We detected a significant decrease in the slope's first resonance frequency ($-24$\%) during the 6 months preceding complete collapse. We also revealed details of slope damage processes acting within the slope, showing sudden breakage phases alternating with smoother deformation and sliding phases. These processes would not have been suspected with slope deformation monitoring alone. The use of a new parameter that combines passive seismic tracking and surface deformation measurements could help revealing active slopes with ongoing damage processes. This innovative approach represents an asset for future practical rock slope monitoring.

\section{Introduction}

Warning systems are of critical importance for unstable rock slope risk mitigation (Sättele et al., 2016). Most forecasting systems rely on ground surface deformation monitoring, assuming that the movements reflect overall slope stability (Intrieri et al., 2019; Lacasse et al., 2009). Ground motion extrapolation is then used to predict the time of slope failure (Hutchinson, 2001; Tavenas \& Leroueil, 1981). Such forecasts have been successful in some cases (Azimi et al., 1988; Krähenbühl, 2006; Wegmann et al., 2003) but also misleading when ambiguous deformation occurs before rupture (Arosio et al., 2009; Rose \& Hungr, 2007; Rosser et al., 2007). Furthermore, slope deformation is not a direct proxy of preparatory damage processes acting within the rock mass (Lacasse et al., 2008; Loew et al., 2012).

Structural health monitoring techniques give access to global dynamic parameters of the structure (Brincker et al., 2001; Fu \& He, 2001). These techniques are commonly used to detect, locate and quantify damage (Ivanović et al., 2000; Lorenzoni et al., 2018). In recent years, structural health monitoring based on ambient vibrations has been extensively applied to rock slopes and structures (e.g., Bessette-Kirton et al., 2022; Finnegan et al., 2022; Geimer et al., 2020; Häusler et al., 2019, 2021, 2022; Kleinbrod et al., 2017). Permanent, passive monitoring systems remain rare, however, because of cost and instrumentation challenges in remote or steep areas (Manconi et al., 2018). For natural rock structures, interpreting dynamic parameter changes is challenging because of poorly known boundary conditions, heterogeneous and discontinuous material properties, and strong environmental forcing (Bottelin, Lévy, et al., 2013; Colombero et al., 2021; Geimer et al., 2022; Kleinbrod et al., 2019). Despite these restrictions, past long-term monitoring of a 21,000~m$^3$ limestone rock column revealed a 30\% decrease in resonance frequency a few weeks before its collapse (Lévy et al., 2010). The resonance frequency decrease was attributed primarily to a reduction in contact stiffness between the column and adjacent stable rock mass due to rock bridge breakage. However, the identification and the monitoring of preparatory damage processes was complicated by instrumentation gaps, notably close to the failure. Reciprocally, an increase in resonance frequencies was attributed to the additional stiffness provided by rock bolts during the monitoring of cliff reinforcement works (Bottelin et al., 2017). However, these past case studies were restricted to monolithic, column-like structures with a major decoupling fracture. The feasibility of passive monitoring for unstable slopes with unclear geometry and complex fracture pattern hence remained to be proven.

In this article, we present the passive seismic monitoring of a $\sim$14,000~m$^3$ complex unstable slope spanning over 10 years. We found that $\fz$ shows clear, unequivocal changes in the last 6 months prior to rockfall, compared to previous baseline value. Passive monitoring revealed damage processes that would not have been detected with slope deformation monitoring only, hence offering additional benefit for risk mitigation.

\section{Site Description}
\label{sec:site}

\begin{figure}[htbp]
\centering
\includegraphics[width=\textwidth]{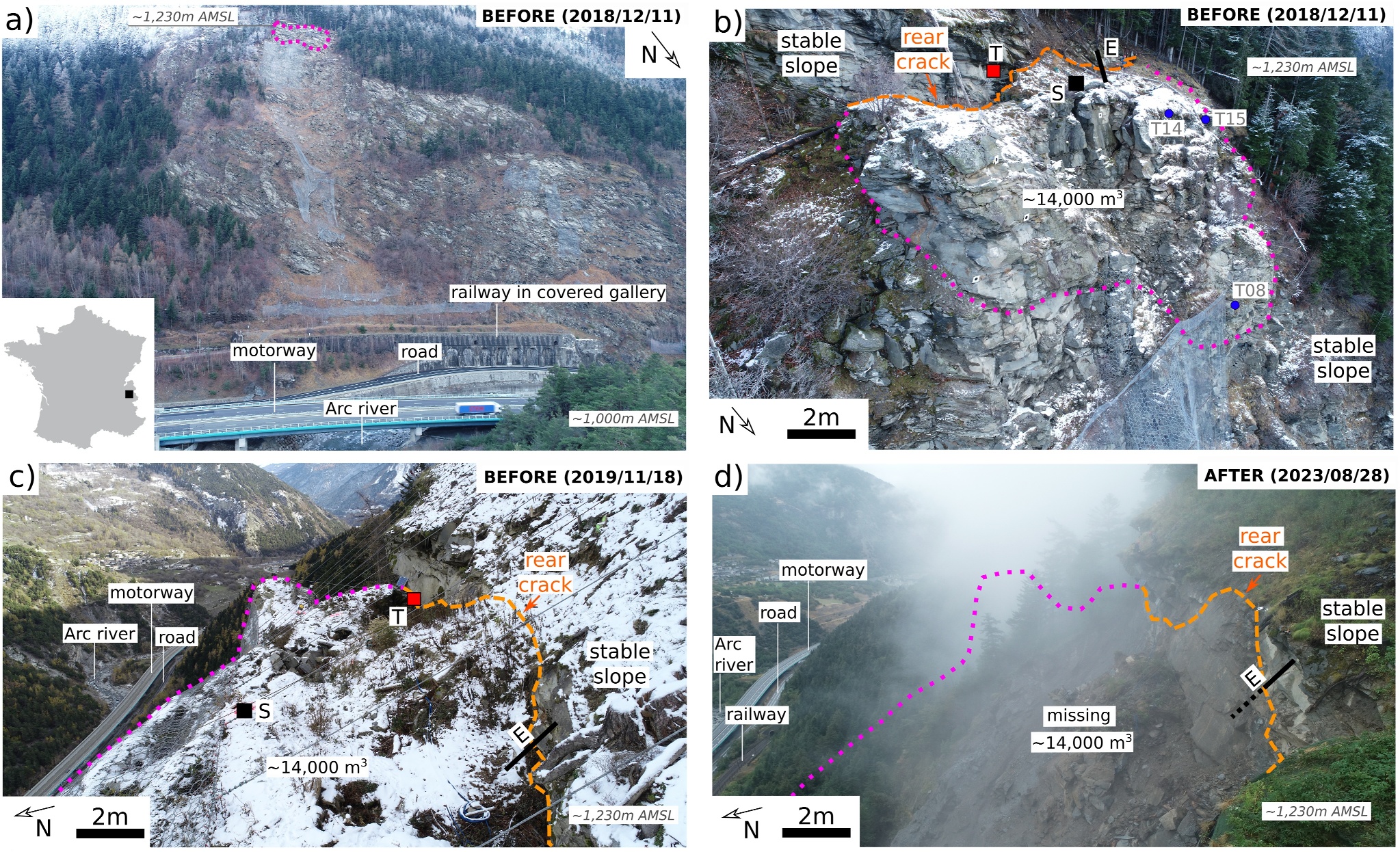}
\caption{General (a) and detailed (b) view of the La Praz site before failure. Side view before (c) and after (d) failure. The $\sim$14,000~m$^3$ unstable volume (purple dotted line) is separated from nearby stable rock slope by a main rear crack (orange dashed line). The extensometer E is set up across the rear crack (thick black segment in b, c, d). Total station monitoring targets are shown with blue circles in (b). The seismic sensor (S, black square) and air temperature datalogger (T, red square) are shown in (b, c). Inset location map is shown in (a).}
\label{fig:site}
\end{figure}

The La Praz site is located at about 1,230~m AMSL on a northern facing flank (Figures~\ref{fig:site}a and \ref{fig:site}b). It is composed of heavily fractured sandstone and shale beds forming an unstable $\sim$14,000~m$^3$ rock mass. The complex and heterogeneous rupture mechanism combines settlement, toppling and swelling (Bottelin, Jongmans, et al., 2013). A 5--6~m vertical main rear crack (Figures~\ref{fig:site}b and \ref{fig:site}c) delineates the unstable slope from nearby stable mass. The La Praz slope is of major importance because of infrastructure assets exposed about 250~m below (see Figures~\ref{fig:site}a and \ref{fig:site}c). The site has been closely monitored since 2006, notably by the regional road operator. Protection works have been stepped up since 2018--2019. The entire $\sim$14,000~m$^3$ unstable slope ultimately collapsed on the 27-08-2023 (see Figures~\ref{fig:site}c and \ref{fig:site}d). The basal rupture surface observed after the collapse was nearly parallel to the slope. This major rockfall caused a 2-week long interruption of the France-Italy motorway, while both the railway and the regional road remain closed at time of writing.

\section{Materials and Methods}

\subsection{Weather Data}

Air temperature was measured directly on site (in a shady area, Figures~\ref{fig:site}b and \ref{fig:site}c) by a PT100 probe connected to a Paratronic\textsuperscript{\textregistered} datalogger. Rainfall time-series were assembled from four nearby stations because of operating gaps (see Table S1 in Supporting Information S1).

\subsection{Slope Deformation Monitoring}

Slope deformation was monitored via both main rear crack aperture tracking with invar wire extensometers (2006--2023) and cliff motion tracking using an automatic total station (2018--2023). The extensometer E (Figures~\ref{fig:site}b--\ref{fig:site}d) and targets T08, T14, T15 (Figure~\ref{fig:site}b) were selected for analysis because they were representative of the slope's overall deformation and showed good measurement continuity. Extensometer aperture is shown in Figure~\ref{fig:monitoring}d. The instrument broke 7 days prior to the collapse due to excessive aperture (more than 1.3~m opening). Total station measurement curve used for 2023 analysis (Figure~\ref{fig:monitoring}h) corresponds to the mean 3D displacement of targets T08, T14, T15. The last total station measurement was acquired about 20~min prior to collapse (15h00 UTC).

\subsection{Passive Seismic Data}

Seismic data were acquired during three separate instrumentation phases (P-01 to P-03, Table S2 in Supporting Information S1). The continuous data streaming phase (P-01, 2010--2013) later changed to only $2 \times 20$~min/day triggered recording in phase P-03 (year 2023) because of power consumption issues. Lennartz\textsuperscript{\textregistered} sensors and Kinemetrics\textsuperscript{\textregistered} digitizer seismological instruments used in phase P-01 (see location in Figure~\ref{fig:site}) were replaced by a low-cost, Raspberry-Shake\textsuperscript{\textregistered} alternative in phase P-03. Analysis was performed using obspy (Beyreuther et al., 2010; Krischer et al., 2015) and scipy Python packages. Signal power spectral density (PSD) was computed using Welch's algorithm with 30~s long tapered windows, no overlapping and Konno-Ohmachi smoothing (Konno \& Ohmachi, 1995). The fundamental resonance frequency $\fz$ of the unstable slope was extracted from PSDs in the [4.0--6.5~Hz] frequency range (see justification in Bottelin, Jongmans, et al., 2013), $\fz$ time-series were sampled with a 12~hr period. The last exploitable $\fz$ measurement dates back from 27-08-2023 00h00 UTC, that is, about 15~hr prior to the slope failure. The seismic instrumentation was lost during the collapse (Figure~\ref{fig:site}d).

\section{Results}

\subsection{Deformation Monitoring}
\label{sec:deformation}

\begin{sidewaysfigure}
\centering
\includegraphics[width=\textheight]{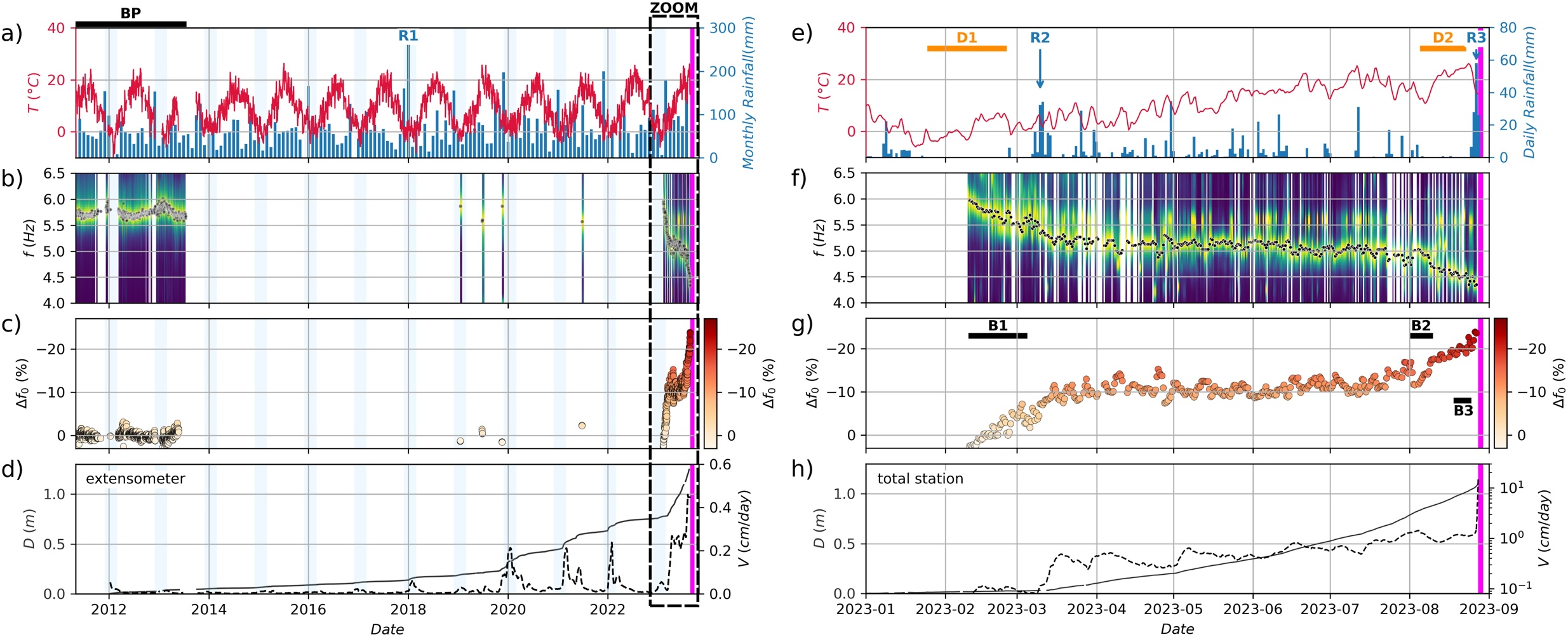}
\caption{Monitoring of the La Praz site. (a--d) Period 2011--2023 and (e--h) zoom on 2023 only. (a, e) Air temperature (solid red line) and rainfall (blue bars), along with rainfall episodes (R1, R2, R3) and dry periods (D1 and D2). (b, f) Ambient vibration normalized Power Spectral Density (colored background) and fundamental resonance frequency $\fz$ tracking (black dots). (c, g) Variation in fundamental frequency after temperature correction, $\dfz$ (colored dots). (d, h) Slope deformation D (solid gray line) and deformation rate V (dashed black line). The 27-08-2023 collapse is shown as a vertical magenta bar. The baseline period, BP, is shown at top left for subplots (a--d). Periods B1, B2 and B3 are shown with black rectangles in (g).}
\label{fig:monitoring}
\end{sidewaysfigure}

Long-term slope deformation shows a relatively steady trend from 2012 until 2018, with an overall $\sim$0.15~m crack opening (Figure~\ref{fig:monitoring}d) and a deformation rate of about 2.5~cm/year ($\sim 7 \times 10^{-3}$~cm/day). Deformation velocity shows a seasonal trend with acceleration during winter and deceleration during summer. A clear acceleration occurred at the beginning of 2018 after the event R1 (about 250~mm accumulated rainfall in January, Figure~\ref{fig:monitoring}a). Similar accelerations were observed during the following winters (especially 2019--2020, 2020--2021 and 2021--2022, velocity peaks close to $\sim$0.2~cm/day), but velocity decreased rapidly during following spring and summer periods. Slope activity then changed at the beginning of 2023. A sharp rise in velocity ($\sim$0.3--0.4~cm/day) occurred almost immediately after rainfall event R2 in early March 2023 (120~mm in 6 days, Figure~\ref{fig:monitoring}e). However, slope deformation rate then went on increasing from $\sim$0.3 to $\sim$1.5~cm/day between mid-March and late-August 2023 (bumpy curve, Figures~\ref{fig:monitoring}d and \ref{fig:monitoring}h). A slight deceleration was observed at the beginning of the dry period D2, yet followed by a clear acceleration during rainfall event R3 (25-08-2023 until 28-08-2023, 120~mm cumulated rainfall in 4 days). The ultimate acceleration was particularly sudden: velocity skyrocketed from about 1.5~cm/day at 00h00 UTC on the 27-08-2023 to about 20~cm/day 15~hr later. The slope collapse started at about 15h20 UTC (at least 5,000~m$^3$ initial rockfall) but later failure sequence is poorly known because of bad weather conditions and darkness through the night.

\subsection{Passive Seismic Monitoring}

\subsubsection{Long-Term Evolution (2011--2023)}
\label{sec:longterm}

During 2011--2023, the fundamental resonance frequency $\fz$ of the instability wanders between about 5.5 and 6.0~Hz (Figure~\ref{fig:monitoring}b). Reversible seasonal variations of $\pm$3.4\% around the mean (3 standard-deviation interval) are observed, with peaks in winter and lows in summer. Such $\fz$ wander related to environmental factors is commonly observed for rock slopes (Colombero et al., 2021; Geimer et al., 2022; Guillemot et al., 2022; Häusler et al., 2021; Kleinbrod et al., 2019). During this 2011--2013 baseline period, the relationship between air temperature T and $\fz$ is shown in Figure S1a in Supporting Information S1 (gray dots). We observe a narrow distribution of the measurements around the linear regression (black line). This overall negative correlation can be explained by the dependence of rock elastic modulus to temperature (Bottelin, Lévy, et al., 2013; Valentin et al., 2017). We then computed the amount of $\fz$ variation not explained by temperature effects as the $\dfz$ parameter (see details in Figure S1b in Supporting Information S1). $\dfz$ varies by about $\pm$2.4\% during the 2011--2013 baseline period (Figure~\ref{fig:monitoring}c). Interestingly, punctual measurements (1-day long surveys) conducted in 2019 and 2021 (black dots) show $\dfz$ lying within a similar $\pm$3\% range. In the same line, the measurements conducted in February 2023 lie within the same range. In contrast, a dramatic change in the site's seismic properties occurred in March 2023 and after, with a dramatic $-1.6$~Hz drop in $\fz$ (Figures~\ref{fig:monitoring}b and \ref{fig:monitoring}f). This sudden reduction far exceeds the magnitude of the $\fz$ wander observed during the previous baseline period.

\subsubsection{Irreversible Damage and Collapse Stages}

We observe three successive stages in slope's resonance frequency response in the months prior to collapse (Figure~\ref{fig:monitoring}f), $\fz$ shows an initial rapid decrease (February--March), followed by a relatively steady period (April--July), then drops again rapidly during August. As visible in Figure~\ref{fig:monitoring}g, most $\dfz$ values between 10 and 25 February still lie within the $\pm$2.4\% range observed during the baseline. This suggests that negligible (or at least non-detectable) damage occurred between the baseline (2011--2013) and the beginning of 2023. In contrast, $\dfz$ decreases rapidly (0\% to about $-10$\%) between the end of February and mid-March 2023. This decrease started in late February, during a period with positive air temperatures and dry conditions (D1, Figure~\ref{fig:monitoring}g). $\dfz$ went on decreasing during rainfall episode R2 (7--14 March), then dropped from about $-10$\% down to $-15$\% between April and early August. $\dfz$ shows a sharper decrease in early August at the start of dry and very warm period D2 (Figure~\ref{fig:monitoring}g). $\dfz$ continued to decrease afterward and stabilized at about $-20$\% between 16 and 25 August. During the first 2 days of rainfall episode R3 (25--27 August), $\dfz$ decreased gradually from about $-20$\% down to $-24$\% (last measurement taken about 15~hr prior to the collapse).

\section{Discussion}

\subsection{Fundamental Frequency Decrease}

We observed a 24\% decrease in $\fz$ at La Praz prior to slope collapse, after correction from thermal effects (see ultimate $\dfz$ in Figure~\ref{fig:monitoring}g). The magnitude of this change agrees well with previous data: the sole past case study reported a 30\% drop in $\fz$ for a 21,000~m$^3$ toppling limestone rock column a few weeks before collapse (Lévy et al., 2010), without correction for thermal effects. Despite different geology and rupture mechanisms, these two 14,000--21,000~m$^3$ case studies showed similar change in fundamental frequency $\fz$ prior to collapse. This order of magnitude also fits with numerical studies reporting 20\%--60\% decrease in $\fz$ prior to failure (Jensen \& Moore, 2023; Lévy et al., 2010). This suggests that a fundamental frequency decrease of a few tens of percent could reflect major rock slope damage and indicate impending collapse.

\subsection{Fundamental Frequency Versus Slope Deformation}

\begin{figure}[htbp]
\centering
\includegraphics[width=0.6\textwidth]{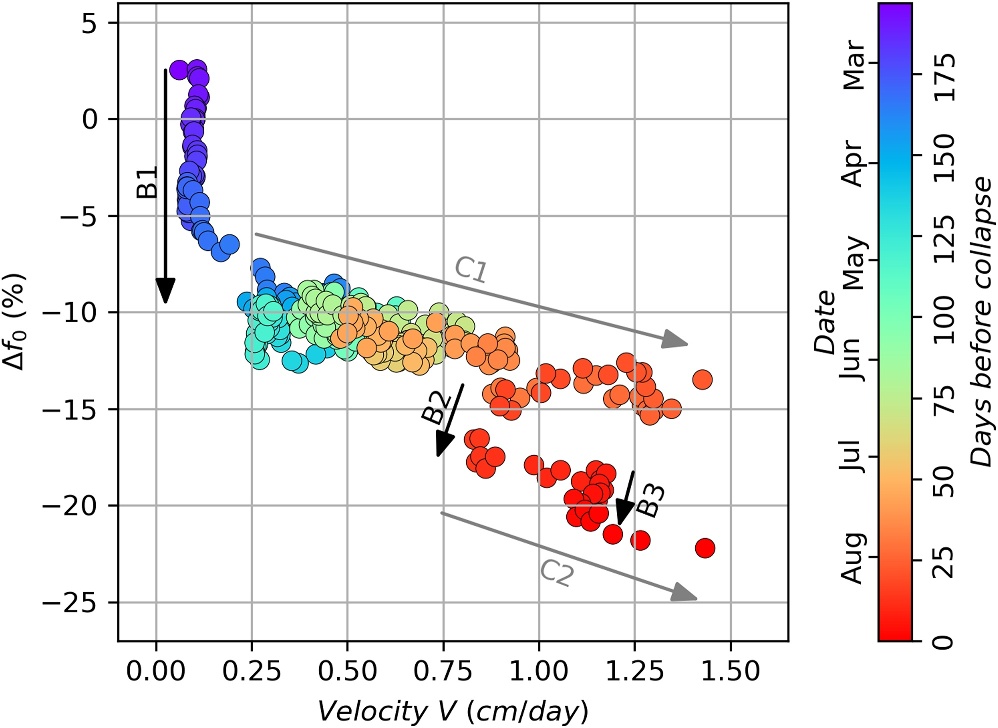}
\caption{$\dfz$ as a function of slope deformation velocity V from total station measurements (period February until 27-08-2023 00h00 UTC). The dot color scales both with the date and the remaining time before collapse (see colorbar). Creep-like (c) or Brittle-failure (b) dominating phases are shown with gray and black arrows, respectively.}
\label{fig:dfv}
\end{figure}

The relationship between $\dfz$ and slope deformation rate is shown in Figure~\ref{fig:dfv}. The final measurement ($V \sim 1.5$~cm/day and $\dfz \sim -24$\%) dates from 00h00 UTC on 27-08-2023, that is, about 15~hr prior to collapse. The variation of $\dfz$ with velocity shows an overall negative correlation (as observed in Figure~\ref{fig:monitoring}). This suggests that both slope deformation and $\dfz$ parameters are closely tied with boundary conditions and bulk rock stiffness, and ultimately reflect the stability conditions of the moving mass. A closer examination, however, reveals that decreases in $\dfz$ may occur without concomitant increase in velocity. For example, the February-early March 2023 $-10$\% drop in $\dfz$ (see B1 in Figures~\ref{fig:monitoring}h and \ref{fig:dfv}) occurred during a slow movement period ($V \le 0.25$~cm/day). Similar sudden $\dfz$ drops were observed again at the beginning and at the end of the dry period D2 (see B2 and B3 in Figures~\ref{fig:monitoring}h and \ref{fig:dfv}). Conversely, notable increases in velocity occurred during winters of 2018, 2020, 2021 and 2022 (Figure~\ref{fig:monitoring}c) without significant ---or at least detectable--- variation in $\dfz$ (still within the $\pm$3\% range). At this closer scale, the relationship between slope deformation and stability conditions hence appears to be more ambiguous than generally assumed in literature (Intrieri et al., 2019).

These results suggest the presence of two different damage processes acting at slope scale. Rapid drops in rock mass stiffness (drops in $\dfz$) without concomitant slope acceleration may be related to brittle failure of key features controlling slope stability (black arrows, Figure~\ref{fig:dfv}), such as intact rock bridges and/or interlocked joint asperities present along the rupture surface (Einstein et al., 1983; Havaej et al., 2015). This was confirmed by numerical simulations (see Figure S2 and Text S1 in Supporting Information S1), where a reduction in stiffness along the rupture surface led to a large drop in $\fz$ (a few tens of percents) without causing significant slope deformation. In contrast, smoother decreases in rock mass stiffness accompanying increasing slope deformation could indicate progressive failure and/or sliding along a pre-existing plane (Section~\ref{sec:site}) and/or ductile, creep-like damage (gray arrows, Figure~\ref{fig:dfv}). These two damage mechanisms are not exclusive at the slope scale: they may act concomitantly and their respective importance could change with time and deformation (Eberhardt et al., 2004).

\subsection{Application to Slope Monitoring}

\begin{figure}[htbp]
\centering
\includegraphics[width=\textwidth]{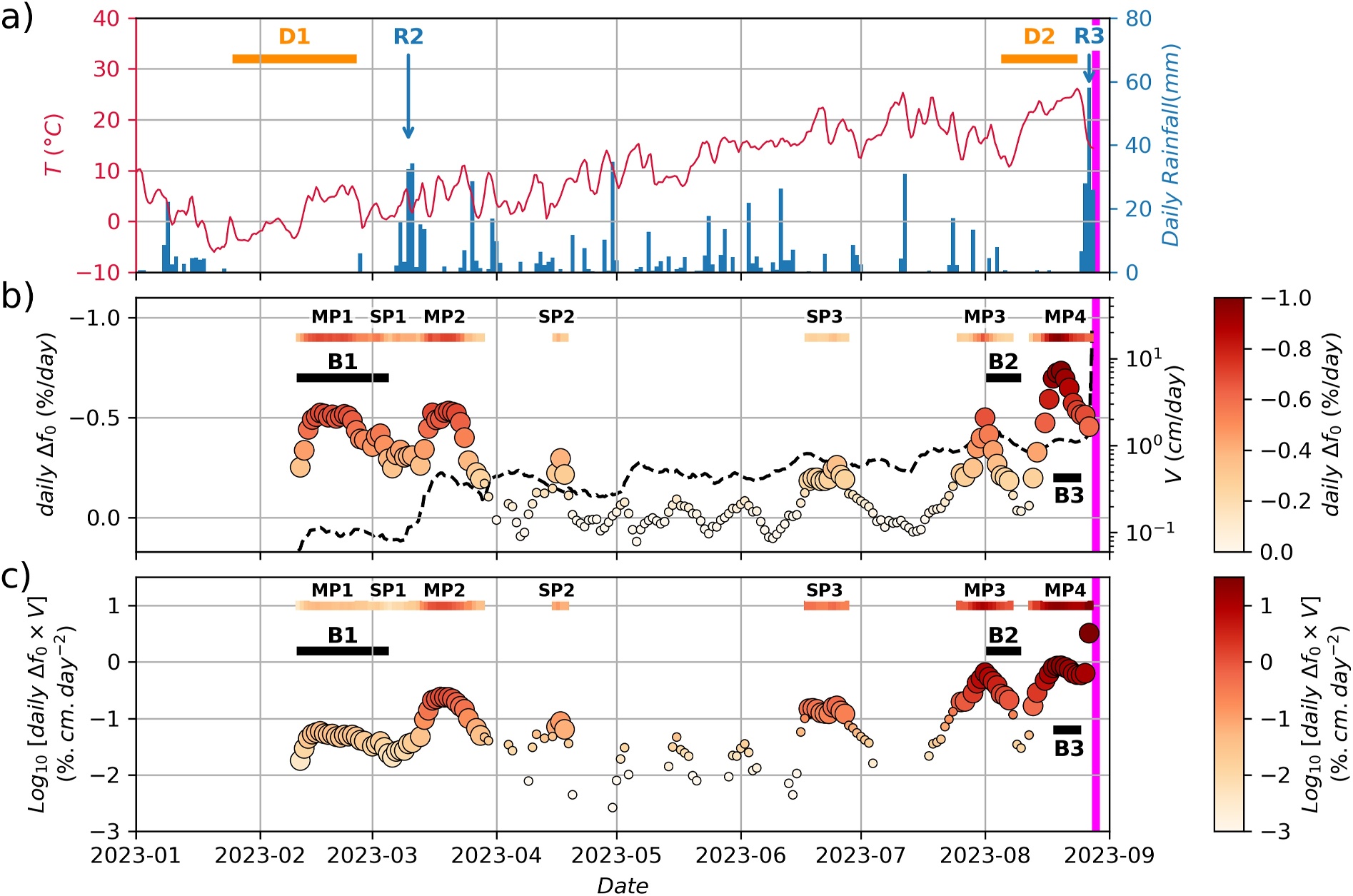}
\caption{(a) Air temperature (red curve) and rainfall (blue bars). Significant rainfall episodes and dry periods are indicated with blue arrows and orange rectangles, respectively. (b) Daily $\dfz$ (colored dots) and deformation rate V (dashed black line). Daily $\dfz$ measurements lying within the baseline mean $\pm$3 standard-deviation interval (i.e., $\pm$0.18\%/day) are shown with smaller dots. (c) Daily $\dfz$ multiplied by deformation rate V (colored dots). Negative values (in case of negative daily $\dfz$) are not represented because of the logarithmic Y-axis. The color for major peaks (MP) and secondary peaks (SP) scales with daily $\dfz$ (b) and daily $\dfz \times V$ (c), respectively. Brittle failure periods revealed in Figure~\ref{fig:dfv} are shown in (b) and (c) as black rectangles.}
\label{fig:daily}
\end{figure}

Tracking the rate of fundamental resonance frequency decrease, for example, at a daily scale, could prove useful for risk mitigation systems. We computed the daily mean variation in $\fz$ after temperature correction both for the 2011--2013 baseline period, and for 2023 only (see Figure S4a, Text S3 in Supporting Information S1). It is referred to as daily $\dfz$ hereafter, and is expressed in \%/day. The daily $\dfz$ shows a narrow and stable normal distribution during the baseline period. In contrast, a strongly skewed distribution (toward negative values) is observed during 2023. The tracking of $\dfz$ time-variations (Figure~\ref{fig:daily}b) shows four main peaks (MP1--MP4, $-0.5$\% to $-0.75$\%/day) and three secondary peaks (SP1--SP3, $-0.25$\%/day). The MP, and to a lesser extent SP, reflect critical periods with intense slope damage. Periods MP2 and MP3 are clearly revealed by both deformation monitoring (rise in velocity, Figure~\ref{fig:daily}b) and daily $\dfz$ (peaks in Figure~\ref{fig:daily}b). In contrast, the major damage period MP1 clearly visible with daily $\dfz$ tracking could not have been detected with slope deformation monitoring nor suspected on the basis of meteorological conditions (Figures~\ref{fig:daily}a and \ref{fig:daily}b). Because ground acceleration in winter was already observed during previous years (see Section~\ref{sec:deformation} and Figure~\ref{fig:monitoring}), air temperature, freeze-thaw alternance, and/or snow melt could be the cause of MP1 (D'Amato et al., 2016; Dochez et al., 2013; Gischig et al., 2011; Krautblatter \& Moser, 2009). This is tempered however by the absence of water (liquid or ice) visible on the La Praz cliff. As for MP1, smaller peaks SP1-SP2-SP3 revealed by daily $\dfz$ are not unambiguously suggested by either meteorological conditions or slope deformation (Figures~\ref{fig:daily}a and \ref{fig:daily}b). We note that brittle damage events B1, B2 and B3 match very well with major peaks in daily $\dfz$ (MP1, MP3 and MP4, respectively, see Figure~\ref{fig:daily}b), restricted to periods with rather small deformation rate (i.e., late MP3 and early MP4).

The MP4 damage period is unequivocally revealed by fundamental frequency tracking, with the sharpest peak in daily $\dfz$ (about $-0.75$\%/day). MP4 begins about 13 days prior to slope collapse, suggesting intense preparatory damage processes at work. Such damage would not have been suspected that early using only deformation monitoring: the velocity showed a rather flat curve during period D2 and until 25-08-2023, that is, 2 days prior to collapse (Figure~\ref{fig:daily}b). Occurring in this context of this intense preparatory damage, the major rainfall event R3 hence appears as the 27-08-2023 collapse triggering factor. Tracking the damage rate acting at the La Praz slope allows to gain new insights in slope's behavior: the jagged, highly variable damage rate observed suggests that the tertiary creep assumption (i.e., increasing deformation occurring under constant stress) was not fulfilled. This unsatisfied condition complicated the use of failure forecasting methods based on slope deformation only (see Figure S3 and Text S2 in Supporting Information S1) and prevented the observation of a linear $\fz$ decrease with time, as suggested by numerical models (Jensen \& Moore, 2023).

In this context, we suggest that both volume and surface deformation monitoring should be combined for improved slope hazard monitoring. We hence propose to multiply the daily variation in fundamental frequency ($\dfz$) by the daily slope deformation velocity ($V$). This new combined parameter scales with acceleration (\%.cm.day$^{-2}$). We show its statistical distribution for both the 2011--2013 baseline period and 2023 (Figure S4b and Text S3 in Supporting Information S1). We observe a very clear separation in distributions between the two periods, and a rise in mean value by about four orders of magnitude. This separation is clearer for the $\dfz \times V$ metric compared to $\dfz$ only. Such marked difference suggests that $\dfz \times V$ values greater than $10^{-3}$\%~cm~day$^{-2}$ could indicate active slopes with ongoing damage. We show the tracking of $\dfz \times V$ during 2023 in Figure~\ref{fig:daily}c. This new metric clearly highlights periods of rapid slope deformation such as the successive major peaks MP2, MP3 and MP4. Interestingly, these successive major peaks show increasing $\dfz \times V$ peak values until failure. As expected, this new monitoring parameter appears yet less sensitive to damage periods without significant slope deformation (B1, B2, B3, Figure~\ref{fig:daily}c), compared to daily $\dfz$ tracking. Based on these results, we believe that this combined, $\dfz \times V$ parameter could help risk mitigation operators in establishing vigilance levels and monitoring unstable slopes.

\subsection{Reversible Environmental Fluctuations}

Environmentally-driven fluctuations in modal parameters have been claimed to complicate stability monitoring of natural structures (Bottelin, Lévy, et al., 2013; Colombero et al., 2018). We overcame this issue by correcting the $\fz$ time-series from the influence of air temperature (see Section~\ref{sec:longterm}). We notably benefited from a long baseline period (more than 2 years), relatively small reversible fluctuations in $\fz$ ($\pm$3.4\% at the seasonal scale, Section~\ref{sec:longterm}), a linear T-$\fz$ relationship (Figure S1 in Supporting Information S1), and negligible influence of other environmental factors. Sites with greater magnitude $\fz$ fluctuation and/or more complex patterns (e.g., hysteresis in the T-$\fz$ relationship, icing effects, water saturation effects, solar radiation) may reveal harder to characterize (e.g., Geimer et al., 2022; Guillemot et al., 2022, 2024; Jensen et al., 2024; Starr et al., 2015). We remain optimistic in the future tackling of this problem for example, with machine learning workflows (Flah et al., 2021; Kim \& Song, 2022; Shi et al., 2022) because seasonal and daily variations in $\fz$ reported in the literature for many natural rock slopes lie within a reasonable range (e.g., $\le$15\% in Colombero et al., 2021, 10\%--25\% in Geimer et al., 2022), and because large $\fz$ decreases observed prior to slope failures (a few tens of percents) should be detectable even with flawed correction of environmental effects.

\section{Conclusions}

We presented analysis of passive seismic monitoring data for a complex, highly fractured rock slope performed until failure. A substantial 24\% decrease in fundamental resonance frequency was detected during the last 6 months prior to collapse, reflecting slope preparatory damage. Tracking of $\fz$ allowed new insights into slope damage processes acting at depth. Sudden decreases in slope stiffness could indicate the failure of key features controlling slope stability such as intact rock bridges and/or interlocked joint asperities. In contrast, gradual loss of slope stiffness may reflect sliding along a pre-existing failure plane and/or ductile deformation. Only some of the slope stiffness drops were clearly associated with forcing events such as very warm periods or intense rainfall events. In addition, intense slope damage episodes would not have been systematically detected by deformation monitoring alone. We hence propose a new monitoring parameter combining slope stiffness tracking and surface deformation. These new insights and focus on key damage periods represent crucial assets for risk mitigation operators, with possible new monitoring applications.

\section*{Data Availability Statement}

Data supporting this research are available in Bottelin and Baillet (2024) with a custom data set term because of the sensitive risk mitigation context of La Praz site. Data access is restricted to research purposes only. Data access can be granted upon request to CD73 (a.lescurier@savoie.fr) and the manuscript's corresponding author.

\section*{Acknowledgments}

CD73, VOR and Alcotra MASSA programs funded this survey. Special thanks to D. Jongmans and E. Larose for their contribution. Thanks to A. Lescurier (CD73), D. Beniamine and D. Rouquet (Tétras Libre); J. Bourdat, L. Mingrat, J.-M. Krysiecki, A. Bouchand, O. Méric and L. Lorier (SAGE Ingénierie) for their support and/or fieldwork.

\section*{Author Contributions}

\textbf{Conceptualization:} P. Bottelin, L. Baillet.
\textbf{Data curation:} P. Bottelin.
\textbf{Formal analysis:} P. Bottelin.
\textbf{Funding acquisition:} P. Bottelin, L. Baillet.
\textbf{Investigation:} P. Bottelin, L. Baillet.
\textbf{Methodology:} P. Bottelin.
\textbf{Project administration:} P. Bottelin.
\textbf{Resources:} P. Bottelin.
\textbf{Software:} P. Bottelin.
\textbf{Supervision:} P. Bottelin, L. Baillet.
\textbf{Validation:} L. Baillet.
\textbf{Visualization:} P. Bottelin.
\textbf{Writing -- original draft:} P. Bottelin.
\textbf{Writing -- review \& editing:} L. Baillet.

\section*{References}
\begingroup
\small
\setstretch{1.1}
\setlength{\parindent}{-1.5em}
\setlength{\leftskip}{1.5em}
\setlength{\parskip}{0.3em}

Arosio, D., Longoni, L., Papini, M., Scaioni, M., Zanzi, L., \& Alba, M. (2009). Towards rockfall forecasting through observing deformations and listening to microseismic emissions. \emph{Natural Hazards and Earth System Sciences}, 9(4), 1119--1131. \url{https://doi.org/10.5194/nhess-9-1119-2009}

Azimi, C., Biarez, J., Desvarreux, P., \& Keime, F. (1988). Prévision d'éboulement en terrain gypseux. In \emph{Proceedings of the 5th international symposium on landslides} (pp. 531--536). A.A. Balkema.

Bessette-Kirton, E. K., Moore, J. R., Geimer, P. R., Finnegan, R., Häusler, M., \& Dzubay, A. (2022). Structural characterization of a toppling rock slab from array-based ambient vibration measurements and numerical modal analysis. \emph{Journal of Geophysical Research: Earth Surface}, 127(8), e2022JF006679. \url{https://doi.org/10.1029/2022JF006679}

Beyreuther, M., Barsch, R., Krischer, L., Megies, T., Behr, Y., \& Wassermann, J. (2010). ObsPy: A Python toolbox for seismology. \emph{Seismological Research Letters}, 81(3), 530--533. \url{https://doi.org/10.1785/gssrl.81.3.530}

Bottelin, P., \& Baillet, L. (2024). Replication data for: Original insights into rock slope damage processes until collapse from passive seismic monitoring [Dataset]. Harvard Dataverse, V1. \url{https://doi.org/10.7910/DVN/Z7PC4G}

Bottelin, P., Baillet, L., Larose, E., Jongmans, D., Hantz, D., Brenguier, O., et al. (2017). Monitoring rock reinforcement works with ambient vibrations: La bourne case study (Vercors, France). \emph{Engineering Geology}, 226, 136--145. \url{https://doi.org/10.1016/j.enggeo.2017.06.002}

Bottelin, P., Jongmans, D., Baillet, L., Lebourg, T., Hantz, D., Lévy, C., et al. (2013). Spectral analysis of prone-to-fall rock compartments using ambient vibrations. \emph{Journal of Environmental \& Engineering Geophysics}, 18(4), 205--217. \url{https://doi.org/10.2113/JEEG18.4.205}

Bottelin, P., Lévy, C., Baillet, L., Jongmans, D., \& Gueguen, P. (2013). Modal and thermal analysis of Les Arches unstable rock column (Vercors Massif, French Alps). \emph{Geophysical Journal International}, 194(2), 849--858. \url{https://doi.org/10.1093/gji/ggt046}

Brincker, R., Zhang, L., \& Andersen, P. (2001). Modal identification of output-only systems using frequency domain decomposition. \emph{Smart Materials and Structures}, 10(3), 441--445. \url{https://doi.org/10.1088/0964-1726/10/3/303}

Colombero, C., Baillet, L., Comina, C., Jongmans, D., Larose, E., Valentin, J., \& Vinciguerra, S. (2018). Integration of ambient seismic noise monitoring, displacement and meteorological measurements to infer the temperature-controlled long-term evolution of a complex prone-to-fall cliff. \emph{Geophysical Journal International}, 213(3), 1876--1897. \url{https://doi.org/10.1093/gji/ggy090}

Colombero, C., Jongmans, D., Fiolleau, S., Valentin, J., Baillet, L., \& Bièvre, G. (2021). Seismic noise parameters as indicators of reversible modifications in slope stability: A review. \emph{Surveys in Geophysics}, 42(2), 339--375. \url{https://doi.org/10.1007/s10712-021-09632-w}

D'Amato, J., Hantz, D., Guerin, A., Jaboyedoff, M., Baillet, L., \& Mariscal, A. (2016). Influence of meteorological factors on rockfall occurrence in a middle mountain limestone cliff. \emph{Natural Hazards and Earth System Sciences}, 16(3), 719--735. \url{https://doi.org/10.5194/nhess-16-719-2016}

Dochez, S., Laouafa, F., Franck, C., Guedon, S., Martineau, F., Bost, M., \& D'Amato, J. (2013). Influence of water on rock discontinuities and stability of rock mass. \emph{Procedia Earth and planetary science. Proceedings of the Fourteenth International Symposium on Water-Rock Interaction}, 14 7, 219--222. \url{https://doi.org/10.1016/j.proeps.2013.03.191}

Eberhardt, E., Spillmann, T., Maurer, H., Willenberg, H., Loew, S., \& Stead, D. (2004). The Randa Rockslide Laboratory: Establishing brittle and ductile instability mechanisms using numerical modelling and microseismicity. In A. A. Balkema (Ed.), \emph{Presented at the 9 th international symposium of landslides} (pp. 481--487). \url{https://doi.org/10.1201/b16816-68}

Einstein, H. H., Veneziano, D., Baecher, G. B., \& O'Reilly, K. J. (1983). The effect of discontinuity persistence on rock slope stability. \emph{International Journal of Rock Mechanics and Mining Sciences \& Geomechanics Abstracts}, 20(5), 227--236. \url{https://doi.org/10.1016/0148-9062(83)90003-7}

Finnegan, R., Moore, J. R., Geimer, P. R., Dzubay, A., Bessette-Kirton, E. K., Bodtker, J., \& Vollinger, K. (2022). Ambient vibration modal analysis of natural rock towers and fins. \emph{Seismological Research Letters}, 93(3), 1777--1786. \url{https://doi.org/10.1785/0220210325}

Flah, M., Nunez, I., Ben Chaabene, W., \& Nehdi, M. L. (2021). Machine learning algorithms in civil structural health monitoring: A systematic review. \emph{Archives of Computational Methods in Engineering}, 28(4), 2621--2643. \url{https://doi.org/10.1007/s11831-020-09471-9}

Fu, Z.-F., \& He, J. (2001). \emph{Modal analysis}. Elsevier. \url{https://doi.org/10.1016/B978-0-7506-5079-3.X5000-1}

Geimer, P. R., Finnegan, R., \& Moore, J. R. (2020). Sparse ambient resonance measurements reveal dynamic properties of freestanding rock arches. \emph{Geophysical Research Letters}, 47(9), e2020GL087239. \url{https://doi.org/10.1029/2020GL087239}

Geimer, P. R., Finnegan, R., \& Moore, J. R. (2022). Meteorological controls on reversible resonance changes in natural rock arches. \emph{Journal of Geophysical Research: Earth Surface}, 127(10), e2022JF006734. \url{https://doi.org/10.1029/2022JF006734}

Gischig, V. S., Moore, J. R., Evans, K. F., Amann, F., \& Loew, S. (2011). Thermomechanical forcing of deep rock slope deformation: 2. The Randa rock slope instability. \emph{Journal of Geophysical Research}, 116(F4), F04011. \url{https://doi.org/10.1029/2011JF002007}

Guillemot, A., Audin, L., Larose, É., Baillet, L., Guéguen, P., Jaillet, S., \& Delannoy, J.-J. (2024). A comprehensive seismic monitoring of the pillar threatening the world cultural heritage site Chauvet-Pont d'Arc cave, toward rock damage assessment. \emph{Earth and Space Science}, 11(4), e2023EA003329. \url{https://doi.org/10.1029/2023EA003329}

Guillemot, A., Baillet, L., Larose, E., \& Bottelin, P. (2022). Changes in resonance frequency of rock columns due to thermoelastic effects on a daily scale: Observations, modelling and insights to improve monitoring systems. \emph{Geophysical Journal International}, 231(2), 894--906. \url{https://doi.org/10.1093/gji/ggac216}

Häusler, M., Gischig, V., Thöny, R., Glueer, F., \& Donat, F. (2022). Monitoring the changing seismic site response of a fast-moving rockslide (Brienz/Brinzauls, Switzerland). \emph{Geophysical Journal International}, 229(1), 299--310. \url{https://doi.org/10.1093/gji/ggab473}

Häusler, M., Michel, C., Burjánek, J., \& Fäh, D. (2019). Fracture network imaging on rock slope instabilities using resonance mode analysis. \emph{Geophysical Research Letters}, 46(12), 6497--6506. \url{https://doi.org/10.1029/2019GL083201}

Häusler, M., Michel, C., Burjánek, J., \& Fäh, D. (2021). Monitoring the preonzo rock slope instability using resonance mode analysis. \emph{Journal of Geophysical Research: Earth Surface}, 126(4), e2020JF005709. \url{https://doi.org/10.1029/2020JF005709}

Havaej, M., Wolter, A., \& Stead, D. (2015). The possible role of brittle rock fracture in the 1963 Vajont Slide, Italy. \emph{International Journal of Rock Mechanics and Mining Sciences}, 78, 319--330. \url{https://doi.org/10.1016/j.ijrmms.2015.06.008}

Hutchinson, J. N. (2001). Landslide risk---To know, to foresee, to prevent. \emph{Geologia Tecnica and Ambientale, Rivista trimestrale dell'Ordine Nazionale del Geologi}, 9, 3--22.

Intrieri, E., Carlà, T., \& Gigli, G. (2019). Forecasting the time of failure of landslides at slope-scale: A literature review. \emph{Earth-Science Reviews}, 193, 333--349. \url{https://doi.org/10.1016/j.earscirev.2019.03.019}

Ivanović, S. S., Trifunac, M. D., Novikova, E. I., Gladkov, A. A., \& Todorovska, M. I. (2000). Ambient vibration tests of a seven-story reinforced concrete building in Van Nuys, California, damaged by the 1994 Northridge earthquake. \emph{Soil Dynamics and Earthquake Engineering}, 19(6), 391--411. \url{https://doi.org/10.1016/S0267-7261(00)00025-7}

Jensen, E. K., \& Moore, J. R. (2023). Coevolution of rock slope instability damage and resonance frequencies from distinct-element modeling. \emph{Journal of Geophysical Research: Earth Surface}, 128(11), e2023JF007305. \url{https://doi.org/10.1029/2023JF007305}

Jensen, E. K., Moore, J. R., Geimer, P. R., \& Finnegan, R. (2024). Combined ambient vibration and surface displacement measurements for improved progressive failure monitoring at a toppling rock slab in Utah, USA. \emph{Frontiers in Earth Science}, 12, 1364653. \url{https://doi.org/10.3389/feart.2024.1364653}

Kim, M., \& Song, J. (2022). Near-Real-time identification of seismic damage using unsupervised deep neural network. \emph{Journal of Engineering Mechanics}, 148(3), 04022006. \url{https://doi.org/10.1061/(ASCE)EM.1943-7889.0002066}

Kleinbrod, U., Burjánek, J., \& Fäh, D. (2019). Ambient vibration classification of unstable rock slopes: A systematic approach. \emph{Engineering Geology}, 249, 198--217. \url{https://doi.org/10.1016/j.enggeo.2018.12.012}

Kleinbrod, U., Burjánek, J., Hugentobler, M., Amann, F., \& Fäh, D. (2017). A comparative study on seismic response of two unstable rock slopes within same tectonic setting but different activity level. \emph{Geophysical Journal International}, 211(3), 1428--1448. \url{https://doi.org/10.1093/gji/ggx376}

Konno, K., \& Ohmachi, T. (1995). Smoothing function suitable for estimation of amplification factor of the surface ground from microtremor and its application. \emph{Doboku Gakkai Ronbunshu}, 525, 247--259.

Krähenbühl, R. (2006). Der Felssturz, der sich auf die Stunde genau ankündigte. \emph{Bull. angew. Geol.}, 11, 49--63. \url{https://doi.org/10.5169/seals-226161}

Krautblatter, M., \& Moser, M. (2009). A nonlinear model coupling rockfall and rainfall intensity based on a four year measurement in a high Alpine rock wall (Reintal, German Alps). \emph{Natural Hazards and Earth System Sciences}, 9(4), 1425--1432. \url{https://doi.org/10.5194/nhess-9-1425-2009}

Krischer, L., Megies, T., Barsch, R., Beyreuther, M., Lecocq, T., Caudron, C., \& Wassermann, J. (2015). ObsPy: A bridge for seismology into the scientific Python ecosystem. \emph{Computational Science \& Discovery}, 8(1), 014003. \url{https://doi.org/10.1088/1749-4699/8/1/014003}

Lacasse, S., Eidsvig, U., Nadim, F., Høeg, K., \& Blikra, L. H. (2008). Event tree analysis of Aknes rock slide hazard. In \emph{Proceedings of the 17th international conference on soil mechanics and geotechnical engineering (volumes 1, 2, 3 and 4)}. Presented at the 17th international conference on soil mechanics and geotechnical engineering, Alexandria, Egypt. \url{https://doi.org/10.3233/978-1-60750-031-5-2633}

Lacasse, S., \& Nadim, F. (2009). Landslide risk assessment and mitigation strategy. In K. Sassa \& P. Canuti (Eds.), \emph{Landslides -- disaster risk reduction} (pp. 31--61). Springer. \url{https://doi.org/10.1007/978-3-540-69970-5_3}

Lévy, C., Baillet, L., Jongmans, D., Mourot, P., \& Hantz, D. (2010). Dynamic response of the Chamousset rock column (Western Alps, France). \emph{Journal of Geophysical Research}, 115(F4), F04043. \url{https://doi.org/10.1029/2009JF001606}

Loew, S., Gischig, V., Moore, J., \& Keller-Signer, A. (2012). Monitoring of potentially catastrophic rockslides, (Eds.). In \emph{Landslides and engineered slopes: Protecting society through improved understanding}. Presented at the 2nd North Am. Symp. On Landslides and engineered slopes, Berhardt (pp. 101--116). Taylor \& Francis Group, London, Banff, Canada.

Lorenzoni, F., Caldon, M., da Porto, F., Modena, C., \& Aoki, T. (2018). Post-earthquake controls and damage detection through structural health monitoring: Applications in l'Aquila. \emph{J Civil Struct Health Monit}, 8(2), 217--236. \url{https://doi.org/10.1007/s13349-018-0270-y}

Manconi, A., Coviello, V., Galletti, M., \& Seifert, R. (2018). Short communication: Monitoring rockfalls with the Raspberry shake. \emph{Earth Surface Dynamics}, 6(4), 1219--1227. \url{https://doi.org/10.5194/esurf-6-1219-2018}

Rose, N. D., \& Hungr, O. (2007). Forecasting potential rock slope failure in open pit mines using the inverse-velocity method. \emph{International Journal of Rock Mechanics and Mining Sciences}, 44(2), 308--320. \url{https://doi.org/10.1016/j.ijrmms.2006.07.014}

Rosser, N., Lim, M., Petley, D., Dunning, S., \& Allison, R. (2007). Patterns of precursory rockfall prior to slope failure. \emph{Journal of Geophysical Research}, 112(F4), F04014. \url{https://doi.org/10.1029/2006JF000642}

Sättele, M., Krautblatter, M., Bründl, M., \& Straub, D. (2016). Forecasting rock slope failure: How reliable and effective are warning systems? \emph{Landslides}, 13(4), 737--750. \url{https://doi.org/10.1007/s10346-015-0605-2}

Shi, S., Du, D., Mercan, O., Kalkan, E., \& Wang, S. (2022). A novel unsupervised real-time damage detection method for structural health monitoring using machine learning. \emph{Structural Control and Health Monitoring}, 29(10), e3042. \url{https://doi.org/10.1002/stc.3042}

Starr, A. M., Moore, J. R., \& Thorne, M. S. (2015). Ambient resonance of mesa Arch, Canyonlands National Park, Utah. \emph{Geophysical Research Letters}, 42(16), 6696--6702. \url{https://doi.org/10.1002/2015GL064917}

Tavenas, F., \& Leroueil, S. (1981). Creep and failure of slopes in clays. \emph{Canadian Geotechnical Journal}, 18(1), 106--120. \url{https://doi.org/10.1139/t81-010}

Valentin, J., Capron, A., Jongmans, D., Baillet, L., Bottelin, P., Donze, F., et al. (2017). The dynamic response of prone-to-fall columns to ambient vibrations: Comparison between measurements and numerical modelling. \emph{Geophysical Journal International}, 208(2), 1058--1076. \url{https://doi.org/10.1093/gji/ggw440}

Wegmann, M., Funk, M., Flotron, A., \& Keusen, H. (2003). Movement studies to forecast the time of breaking off of ice and rock masses. In J. Zschau \& A. Küppers (Eds.), \emph{Early warning systems for natural disaster reduction} (pp. 565--568). Springer. \url{https://doi.org/10.1007/978-3-642-55903-7_76}

\subsection*{References From the Supporting Information}

Fukuzono, T. (1985). A new method for predicting the failure time of a slope. In \emph{Proc. IV international conference and field workshop on landslides}.

Comsol Multiphysics. (2023). Comsol Multiphysics\textsuperscript{\textregistered} v.6.1.

\endgroup

\end{document}